%% file: Dalitz-ICHEP-2004-Paper.tex
\documentclass[11pt]{article}
\usepackage{graphicx}
\usepackage{rotating}

\newcommand{\BABARPubYear}    {04}

\newcommand{\BABARConfNumber} {14}
\newcommand{\SLACPubNumber} {10601}

\input pubboard/babarsym

\input symbols

\long\def\inst#1{\par\nobreak\kern 4pt\nobreak
    {\it #1}\par\vskip 10pt plus 3pt minus 3pt}

\begin{document}


{\pagestyle{empty}

\begin{flushright}
\babar-CONF-\BABARPubYear/\BABARConfNumber \\
SLAC-PUB-\SLACPubNumber \\
July 2004 \\
\end{flushright}

\par\vskip 5cm

\begin{center}
\Large \bf \boldmath
Amplitude Analysis of \BtoPPP\ and \BtoKPP
\end{center}
\bigskip

\begin{center}
\large The \babar\ Collaboration\\
\mbox{ }\\
\today
\end{center}
\bigskip \bigskip

\begin{center}
\large \bf Abstract
\end{center}

We present preliminary results of a maximum-likelihood Dalitz-plot
analysis of the charmless hadronic \Bpm\ decays to the final states
\PPP\ and \KPP\ from data corresponding to an
integrated on-resonance luminosity of \onreslumi\ recorded by the \babar\
experiment at the SLAC PEP-II asymmetric-energy \B\ Factory. We measure
the total branching fractions ${\cal{B}}(\BtoPPP) = (16.2 \pm 2.1 \pm 1.3) \times 10^{-6}$ 
and ${\cal{B}}(\BtoKPP) = (61.4 \pm 2.4 \pm 4.5) \times 10^{-6}$, 
and provide fit fractions and phases for intermediate resonance states.

\vfill
\begin{center}

Submitted to the 32$^{\rm nd}$ International Conference on High-Energy Physics, ICHEP 04,\\
16 August---22 August 2004, Beijing, China

\end{center}

\vspace{1.0cm}
\begin{center}
{\em Stanford Linear Accelerator Center, Stanford University, 
Stanford, CA 94309} \\ \vspace{0.1cm}\hrule\vspace{0.1cm}
Work supported in part by Department of Energy contract DE-AC03-76SF00515.
\end{center}

\newpage
}

\input pubboard/authors_sum2004.tex

\setcounter{footnote}{0}

\section{Introduction}
\label{sec:Introduction}

The study of charmless hadronic \B\ decays can make important
contributions to the understanding of CP violation in the Standard
Model as well as to models of hadronic decays. The \B-meson decay to
the three-body final state can proceed via intermediate resonances formed
from two of the particles. 
These two-body states can interfere with each other and with the nonresonant three-body decay. 
The three-body state is unique in the search for weak phases as it is
possible to determine the strong phase variation for overlapping
resonances.
Studies of these decays can also help to clarify the nature
of the resonances involved, not all of which are well understood.  
A full Dalitz-plot analysis is necessary to correctly model this interference and extract branching fractions. 

Observations of \B-meson decays to the \PPP\ and
\KPP\ three-body final states have already
been reported by the Belle and \babar\ collaborations using a method
that treats each intermediate decay incoherently~\cite{Belle,Babar1, Belle2}. 
Belle has also reported a preliminary Dalitz-plot
analysis of the decay \BtoKPP~\cite{belleDalitz}.
We present a preliminary Dalitz-plot analysis for both the \BtoPPP\ and \BtoKPP\ decay modes.

\input{detector.tex}

\input{selection.tex}

\input{background.tex}

\input{final_data.tex}

\input{amplitude.tex}

\section{Physics Results}
\label{sec:Physics}

\input{results_pipipi.tex}

\input{results_kpipi.tex}

\input{systematics.tex}

\input{summary.tex}

\section{Acknowledgments}
\label{sec:Acknowledgments}

\input pubboard/acknowledgements

\end{document}

%% file: symbols.tex
\newcommand{\NLL}     {\mbox{$-\ln{\cal L}$}}

\def\D       {\ensuremath{D}\xspace}

\newcommand{\onreslumi}  {\mbox{166 \invfb}}
\newcommand{\offreslumi} {\mbox{16 \invfb}}
\newcommand{\bbpairs}    {\mbox{182 million}}

\newcommand{\costtb}     {\mbox{$\cos{\theta_{T}}$}}
\newcommand{\abscosttb}  {\mbox{$\left|\cos{\theta_{T}}\right|$}}

\newcommand{\ppppos}            {\mbox{$\pip  \pim  \pip$}}
\newcommand{\Kpppos}            {\mbox{$\Kp   \pim  \pip$}}

\newcommand{\Btoppppos}         {\mbox{$\Bp \to \ppppos$}}
\newcommand{\BtoKpppos}         {\mbox{$\Bp \to \Kpppos$}}

\newcommand{\PPP}               {\mbox{$\pipm \pimp \pipm$}}
\newcommand{\KPP}               {\mbox{$\Kpm  \pimp \pipm$}}

\newcommand{\BtoPPP}            {\mbox{$\Bpm \to \PPP$}}
\newcommand{\BtoKPP}            {\mbox{$\Bpm \to \KPP$}}

\newcommand{\KstarI}            {\mbox{$\Kstarz(892)$}}

\newcommand{\KstarII}       {\mbox{$\Kstarz_0(1430)$}}

\newcommand{\KstarIII}      {\mbox{$\Kstarz_2(1430)$}}

\newcommand{\KstarIV}       {\mbox{$\Kstarz(1680)$}}

\newcommand{\rhoz}          {\mbox{$\rho^0$}}
\newcommand{\rhop}          {\mbox{$\rho^+$}}

\newcommand{\rhoI}         {\mbox{$\rhoz(770)$}}
\newcommand{\rhoII}         {\mbox{$\rhoz(1450)$}}

\newcommand{\fz}            {\mbox{$f_0(980)$}}

\newcommand{\fzII}          {\mbox{$f_2(1270)$}}

\newcommand{\fzIII}         {\mbox{$f_0(1370)$}}

\newcommand{\Dzbpi}         {\mbox{$\Dzb \pip$}}

\newcommand{\BtoDzbpi}      {\mbox{$B^+ \to \Dzbpi$}}

\newcommand{\DzbtoKpi}      {\mbox{$\Dzb \to \Kp\pim$}}

\newcommand{\JPsitoll}   {\mbox{$\jpsi \to \ellell$}}

\newcommand{\Psitoll}   {\mbox{$\psitwos \to \ellell$}}

\newcommand{\BtoetapKp}       {\mbox{$\Bu \to \etapr \Kp$}}

\newcommand{\BztoKpipiz}      {\mbox{$\Bz \to \Kp\pim\piz$}}

\newcommand{\BztorhoK}  {\mbox{$\Bz \to \rhop \Km$}}

\newcommand{\DE}{\ensuremath{\Delta E}}

\def\modeIII{B^{\pm} \rightarrow \rho^0(770) \pi^{\pm}}

\def\Y#1S{{\Upsilon\rm(#1S)}}

\def\beq{\begin{equation}}
\def\eeq{\end{equation}}

%% file: pubboard/authors_sum2004.tex
\begin{center}
\small

The \babar\ Collaboration,
\bigskip

%
B.~Aubert,
R.~Barate,
D.~Boutigny,
F.~Couderc,
J.-M.~Gaillard,
A.~Hicheur,
Y.~Karyotakis,
J.~P.~Lees,
V.~Tisserand,
A.~Zghiche
\inst{Laboratoire de Physique des Particules, F-74941 Annecy-le-Vieux, France }
% ...

\end{center}\newpage

%% file: detector.tex
\section{\boldmath The \babar\ Detector and Data Sample}
\label{sec:babar}

Here we present preliminary results from a full amplitude analysis
based on a \onreslumi\ data sample containing \bbpairs\ \BB\ pairs
collected with the \babar\ detector~\cite{babardet} at the SLAC PEP-II
asymmetric-energy \epem\ storage ring~\cite{pep} operating at the $\Upsilon(4S)$
resonance at a center-of-mass energy of $\sqrt{s}=10.58$\gev.
An additional total integrated luminosity of \offreslumi\ 
was recorded at an energy $40$\mev below this energy and
is used to study backgrounds from continuum production. The charm
decay \BtoDzbpi, \DzbtoKpi\ is used as a calibration
channel as it has a relatively high branching fraction.

Details of the \babar\ detector are described elsewhere~\cite{babardet}. 
The specific components used for this
paper are charged particle tracking provided by a combination of a
silicon vertex tracker (SVT), which consists of five layers of
double-sided detectors, and a 40-layer central drift chamber (DCH) in a
1.5-T solenoidal magnetic field. This allows a transverse momentum
resolution for the combined tracking system of $\sigma_{p_T}/p_T =
0.0013p_T \oplus 0.0045$, where the sum is in quadrature and $p_T$ is
measured in \gevc. Charged-particle identification is provided by
combining information on the average energy loss $(dE/dx)$ in the two
tracking devices and the angle of emission of Cherenkov radiation in an
internally reflecting ring-imaging Cherenkov detector (DIRC) covering
the central region. The $dE/dx$ resolution from the drift chamber is
typically about $7.5 \%$ for pions. The Cherenkov angle resolution of
the DIRC is measured to be 2.4 mrad, which provides nearly $3\sigma$
separation between charged kaons and pions at a momentum of 3 \gevc.

%% file: selection.tex
\section{Event Selection and Reconstruction}

\B-meson candidates are reconstructed from events that have four or more charged tracks.
Each track is required to have at least 12 hits in the DCH, 
a minimum transverse momentum of 100 \mevc, and a distance of closest approach
to the primary vertex of less than 1.5\cm\ in the transverse plane and
less than 10\cm\ along the beam axis. Charged tracks identified as leptons are rejected. 
The \B-meson candidates are formed from three-charged-track
combinations and particle identification criteria are applied. 
The average selection efficiency for kaons in our final state that have passed the tracking
requirements is $\sim80\%$ including geometrical acceptance, while the
misidentification probability of pions as kaons is below 5\% at all momenta.  
The kaon veto on pions in our final state is $\sim98\%$ efficient.
The \B-meson candidates' energies and
momenta are required to satisfy appropriate kinematic
constraints, as detailed in Section~\ref{sec:finalsel}.

%% file: background.tex
\section{Background Suppression and Characterisation}
\label{sec:background}

The dominant source of background comes from light quark and charm
continuum production.  This background is suppressed
by imposing requirements on event-shape variables
calculated in the $\FourS$ rest frame.  The first discriminating
variable is \costtb, the cosine of the angle between the thrust axis 
of the selected \B\ candidate and the thrust axis of the rest of the event.  
For continuum background the distribution of \abscosttb\ is strongly peaked towards
unity whereas the distribution is uniform for signal events.  We
require \abscosttb\ $< 0.575$ for \Btoppppos and \abscosttb\ $< 0.65$ for \BtoKpppos.\footnote{Charge-conjugate states are implied throughout this section.} 

Additionally, we make requirements on a Fisher discriminant $\cal{F}$~\cite{fisher} formed using a linear
combination of nine variables representing the angular distribution of
the energy flow of the rest of the event into each of nine two-sided
concentric 10$^{\circ}$ cones around the thrust axis of the reconstructed \B~\cite{CLEO}.  

Other backgrounds arise from \BB\ events. There
are four main sources: combinatorial background from three
unrelated tracks; three- and four-body $\B \to D X$ decays, where $X$ represents other particles in the final state; charmless
four-body decays with a missing particle and three-body decays with one
or more particles misidentified. In the case of charm decays with
large branching fractions these backgrounds are greatly reduced by
vetoing the appropriate region of the two-body invariant-mass spectra.  
The rejected decays and the invariant-mass veto ranges are given in 
Table~\ref{tab:veto}.

\begin{table}[hbt]
\caption{The invariant-mass veto ranges (in \gevcc) for intermediate resonances for \ppppos\  and \Kpppos.
         The leptons in the \jpsi\ and \psitwos\ decays are misidentified as pions.}
\label{tab:veto}
\begin{center}
\begin{tabular}{|c|c|c|} \hline
Resonance              & \ppppos      & \Kpppos                    \\
\hline
\JPsitoll & $3.05 < m_{\pipi} < 3.14$ & $2.97 < m_{\pipi} < 3.17$   \\
\Psitoll  & $3.64 < m_{\pipi} < 3.73$  & $3.56 < m_{\pipi} < 3.76$ \\
\DzbtoKpi\ (or \pipi) &  $1.70 < m_{\pipi} < 1.93$  & $1.80 < m_{\Kp\pim} < 1.90$\\
\hline
\end{tabular}
\end{center}
\end{table}

The remaining charm backgrounds that escape the vetoes and backgrounds from
charmless \B\ decays are studied using a large sample of Monte Carlo
(MC) simulated \BB\ decays equivalent to approximately five times the
integrated luminosity for the data. 
Any events that pass the selection criteria are further studied using exclusive MC samples 
to estimate reconstruction efficiency and yields. 
The \mes\ and Dalitz distributions of the \BB\ backgrounds, 
which are used in the likelihood fits, 
are then normalised to the total number of predicted \BB\ events in the final data sample.
For \ppppos, we expect $200 \pm 14$ \B-related background events, 
dominated by the decays $\Bp \to \KS\pip$ and \BtoKpppos. 
For \Kpppos, we expect $315 \pm 17$ background events and the dominant 
backgrounds come from \B-meson decays to states containing a 
\D\ or \Dstar\ and a $\rho$ or $\pi$, 
the decays \BztorhoK\ and \BtoetapKp, 
and the nonresonant decay \BztoKpipiz.

A further background in this analysis comes from 
signal events that have been misreconstructed
by switching one or more particles from the decay of the signal
\B\ meson with particles from the other \B\ meson in the event.  
The amount of this background is estimated from MC studies, and for both \Btoppppos\ and
\BtoKpppos\ is found to be a very small effect that accounts for
less than 2\% of the final data sample in the signal box (defined in Section~\ref{sec:finalsel}). As such it is neglected in the analysis.

Both the continuum and \B-related backgrounds are modeled in the Dalitz amplitude fit 
using linearly interpolated 2-dimensional histograms.

%% file: final_data.tex
\section{Final Data Selection}
\label{sec:finalsel}

Two kinematic variables are used to select a final data sample. The
first variable is $\DeltaE = E_B^* - \sqrt{s}/2$, the difference
between the center of mass (CM) energy of the \B-meson candidate and
$\sqrt{s}/2$, where $\sqrt{s}$ is the total CM energy. The second is
the energy-substituted mass \mes\ $= \sqrt{(s/2 + {\bf p_i} \cdot {\bf
p}_B )^2/ E_i^2 - {\bf p}^2_B}$ where {\bf p$_B$} is the $B$ momentum
and (E$_i$,{\bf p$_i$}) is the four-momentum of the initial
state. The mean of the \DeltaE\ distribution is shifted by
$-8.3$\mev\ as measured from the calibration channel \BtoDzbpi. For
\PPP\ we require $ -68.3 < \DeltaE < 51.7$\mev; for \KPP\ the
requirement is $ -38.3 < \DeltaE < 51.7$\mev\ where the lower edge
is tightened by 30\mev\ to reduce \KPP\ specific \BB\ backgrounds.

\begin{figure}[ht]
\begin{center}
\includegraphics[width=0.60\textwidth]{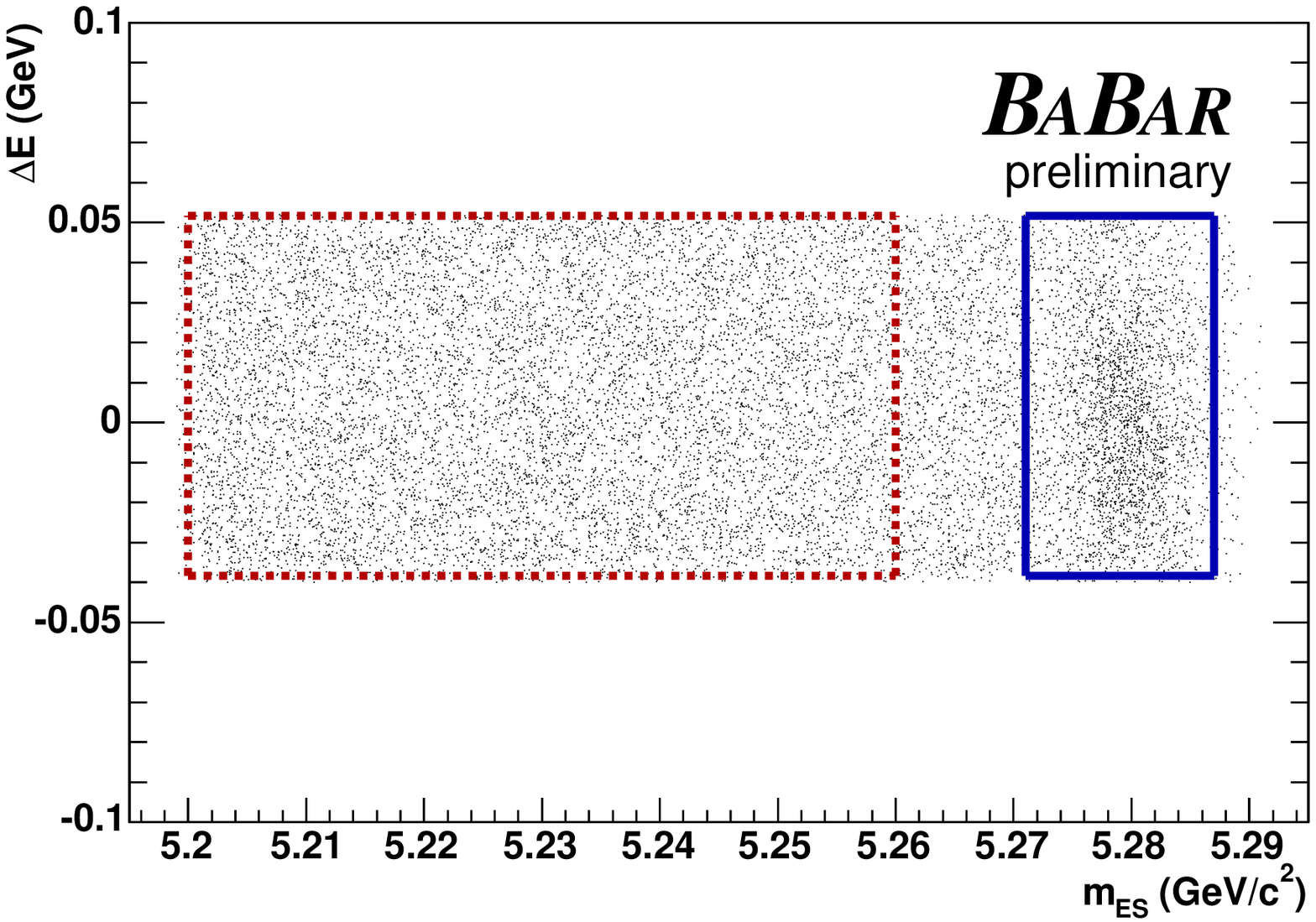}
\caption{\DE--\mes\ plane, showing the signal strip (entire populated region), sideband and signal box regions,
         as defined in the text, for \BtoKPP. Similar regions apply for \BtoPPP.
         The events shown are \BtoKPP\ candidates from the data sample.}
\label{fig:demes}
\end{center}
\end{figure}

We define three regions in the \DeltaE-\mes plane, illustrated in Figure~\ref{fig:demes}.
The signal box is defined by $5.271 < \mes < 5.287$\gevcc\ 
and events in this region are used in the amplitude analysis. 
The signal strip is defined by $5.20 < \mes < 5.29$\gevcc\ 
and is used to determine the fraction of signal and \qqbar\ continuum events in the signal box. 
A sideband area below the signal box, defined by $5.20 < \mes < 5.26$\gevcc, is used to
obtain the distribution of the \qqbar\ continuum events in the Dalitz plane. 

We accept one \B-meson candidate per event in the signal strip. Fewer than 3\% of events have
multiple candidates and in those events one
candidate is randomly accepted to avoid bias.

After all selection criteria are applied the average efficiency for
reconstruction of phase space \PPP\ and \KPP\ MC events in the signal box 
is $(13.00\pm0.04)$\% and $(13.26\pm0.03)$\%, respectively, where the errors are statistical only. 
The efficiency across the Dalitz plot is uniform except
for very small decreases near the boundaries.
In the amplitude analysis, this is taken into
account by calculating the efficiency as a function of position in the
Dalitz plot using phase space MC.

\begin{figure}[!htb]
\begin{center}
\includegraphics[angle=0,width=0.45\textwidth]{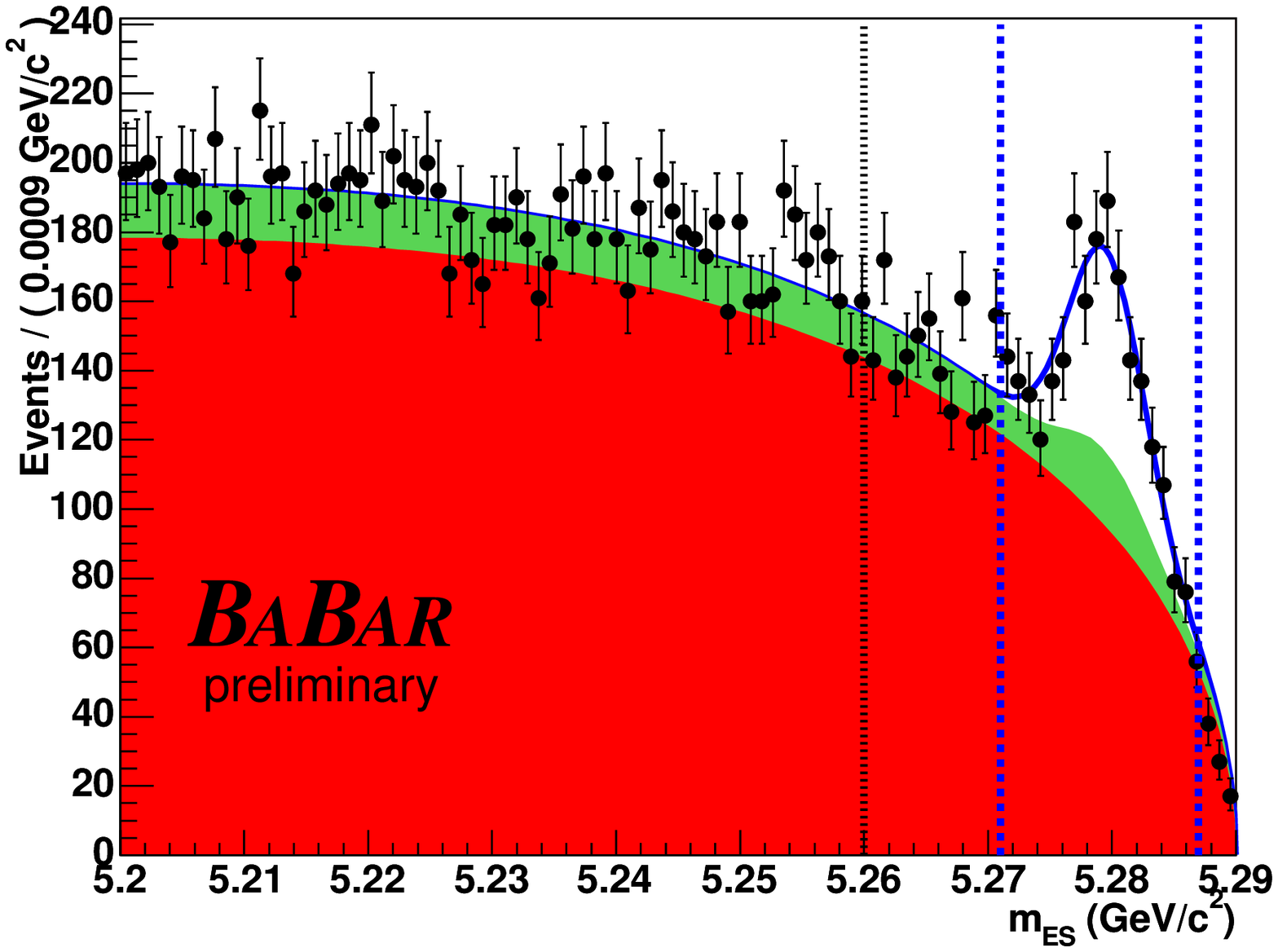}
\includegraphics[width=0.45\textwidth]{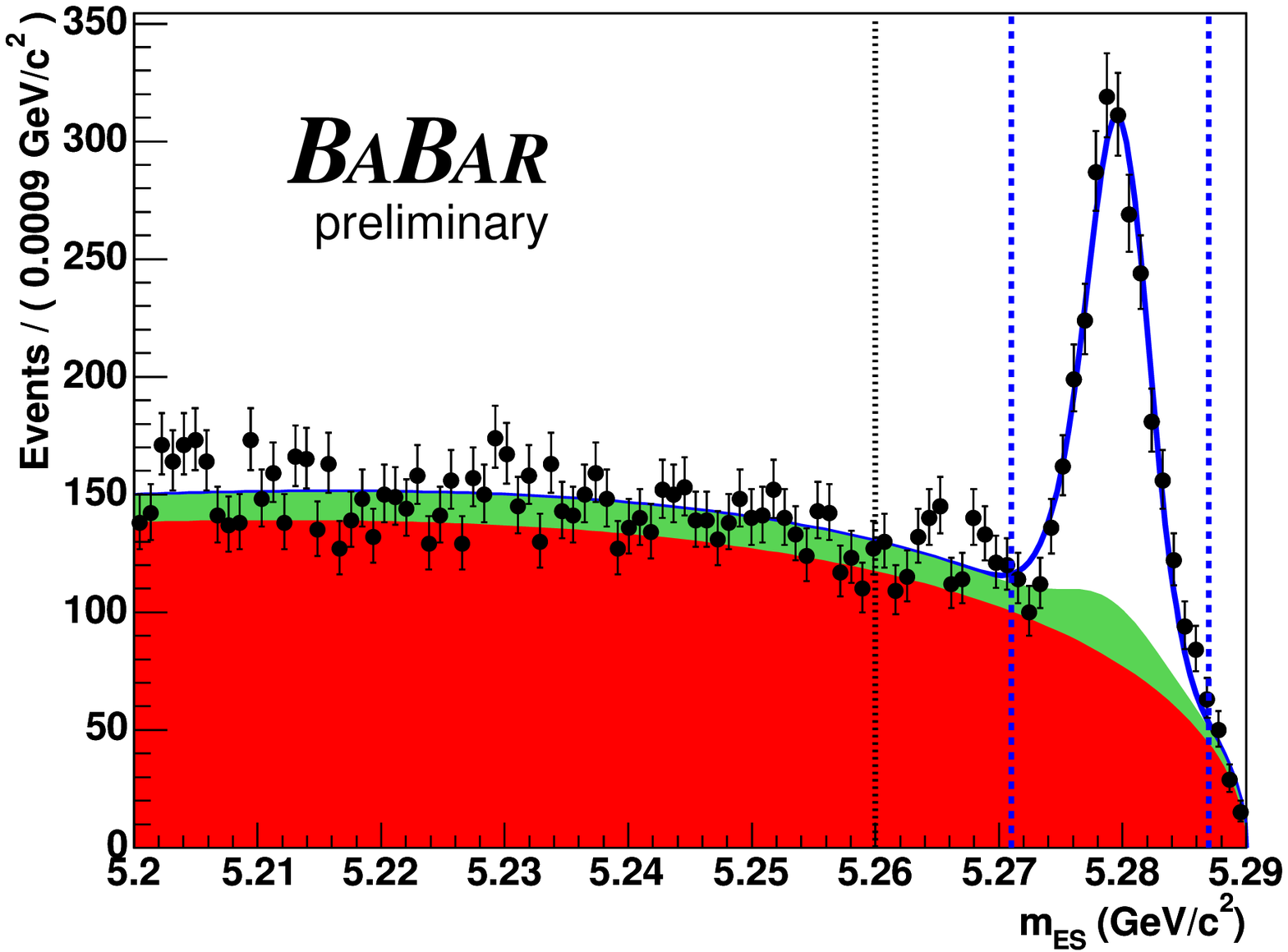}
\caption[\mes\ data and PDFs]
        {\mes\ distribution, together with the fitted PDFs: 
         the data are the black points, 
         the lower solid red area is the \qqbar\ component, 
         the middle solid green area is the \BB\ background contribution, 
         while the upper blue line shows the total fit result. 
         All errors shown are statistical only.
         The left hand plot is for \PPP; the right hand plot is for \KPP.
         The vertical dashed blue lines show the signal region extent, 
         while the dotted black line indicates the upper edge of the sideband.}
\label{fig:mesfit}
\end{center}
\end{figure}

The \mes\ distribution for events in the signal strip is used to determine the
number of \qqbar\ and signal events in the signal box. The signal
component is modeled by a double Gaussian function with parameters
obtained from phase-space MC. These parameters are fixed except for
the mean of the core Gaussian. 
The \qqbar\ continuum background is modeled using the experimentally motivated ARGUS function~\cite{argus} 
with the endpoint fixed to the beam energy while the shape parameter is allowed to float. 
Finally, the \BB\ background is modeled with an ARGUS function
plus a Gaussian to account for peaking \BB\ backgrounds. 
All parameters of the \BB\ component, including the amount of peaking and nonpeaking \BB\ background, are  obtained and fixed from the MC simulation.
The fraction of signal and \qqbar\ events is allowed to float. 
Figure~\ref{fig:mesfit} shows the \mes\ projections of fits to the data for both \PPP\ and \KPP. 
The \chisq\ per degree of freedom for these projections is 0.83 (1.13) for \PPP\ (\KPP).
For \PPP\ the total number of events in the signal box is 2407; 
for \KPP\ the total number of events in the signal box is 3174. 
The extracted fractions of signal, \qqbar\ continuum and \BB\ backgrounds are given
in Table~\ref{tab:fractions}.

\begin{table}[ht]
\caption{Total number of events in the signal box, for \PPP\ and \KPP\ candidates.
Also shown are the fractions for each hypothesis calculated from the fit to the \mes\
distribution for events in the signal strip.
The errors arise from the combination of the fit errors on the signal and \qqbar\ components,
and the error on the \BB\ background due to the uncertainties on the various branching fractions and MC efficiencies.}
\label{tab:fractions}
\begin{center}
\begin{tabular}{|l|c|c|} \hline
 & \PPP             & \KPP \\ \hline
Total Events & 2407  & 3174\\
\hline
 & \multicolumn{2}{|c|}{Fraction of Events}\\ \hline
Signal             & 0.159 $\pm$ 0.021 & 0.467 $\pm$ 0.021\\
\qqbar\ background & 0.758 $\pm$ 0.018 & 0.430 $\pm$ 0.014 \\
\BB\ background    & 0.083 $\pm$ 0.010 & 0.103 $\pm$ 0.006\\ \hline
\end{tabular}
\end{center}
\end{table}

%% file: amplitude.tex
\section{Dalitz Amplitude Analysis}
\label{sec:amplitude}

In terms of a Dalitz-plot analysis of the \B-meson decay to the final state $hhh$ 
(where $h =\pi$ or \kaon) a number of intermediate states contribute and the 
total rate can be represented in the form:

\begin{equation}
{\frac{d\Gamma}{dm_{13}^2 dm_{23}^2}} = |{\cal{M}}|^2
= \left| \sum_i c_i e^{i\theta_i} F_i(m_{13}^2, m_{23}^2) \right|^2
\end{equation}

\noindent where $m_{13}^2$ and $m_{23}^2$ are the invariant masses squared
of pairs of final state particles. 
In the case of \PPP, $m_{13}^2=(p_{\pipm} + p_{\pimp})^2$ and
$m_{23}^2=(p_{\pipm} + p_{\pimp})^2$; 
for \KPP,
$m_{13}^2=(p_{\Kpm} + p_{\pimp})^2$ and $m_{23}^2=(p_{\pipm} +
p_{\pimp})^2$. 
The amplitude for a given decay mode is $c_i e^{i\theta_i} F_i(m_{13}^2, m_{23}^2)$, where
$c_i$ and $\theta_i$ are the unknown real parameters of each partial decay mode,
while $F_i$ describes the dynamics of the amplitudes.
These $F_i$ consist of a product of the invariant mass and angular
distribution probabilities:

\begin{equation}
F_i=R_i(m) \times T_i(\cos\theta_H)
\end{equation}

\noindent where $R_i(m)$ is the resonance mass distribution and $T_i(\cos\theta_H)$ is the angular probability distribution. 
The angle $\theta_H$ is defined as 
the angle between the momentum vector of one of the resonance
daughters in the resonance rest-frame and the momentum vector of the
resonance in the \B\ rest-frame.

To fit the data in the signal box, we define an unbinned likelihood function for one
event to have the form shown in Eq.~\ref{LikeEqn}.
The fit is performed allowing the amplitude magnitudes ($c_i$) and the phases ($\theta_i$) to vary. 

\begin{eqnarray}
\label{LikeEqn}
{\cal{L}}(m_{13}^2,m_{23}^2) & = & (1 - f_{q\bar{q}} - f_{B\bar{B}})~\frac{|\sum_{i=1}^{N} 
c_i e^{i\theta_i} F_i(m_{13}^2,m_{23}^2)|^2\epsilon(m_{13}^2,m_{23}^2)}
{\int~{|\sum_{i=1}^{N} c_i e^{i\theta_i} F_i(m_{13}^2,m_{23}^2)|^2\epsilon(m_{13}^2,m_{23}^2)}~dm_{13}^2dm_{23}^2}\\ \nonumber
& & +~f_{q\bar{q}}~\frac{Q(m_{13}^2,m_{23}^2)}
{\int~Q(m_{13}^2,m_{23}^2)~dm_{13}^2dm_{23}^2}\\ \nonumber
& & +~f_{B\bar{B}}~\frac{B(m_{13}^2,m_{23}^2)}
{\int~B(m_{13}^2,m_{23}^2)~dm_{13}^2dm_{23}^2}
\end{eqnarray}
where

\begin{itemize}
\item $m_{13}^2$ and $m_{23}^2$ are the invariant mass-squared values of the daughter pairs (1,3) and (2,3);
\item $N$ is the number of resonant and nonresonant contributions to the plot;
\item $F_i$ is the dynamical part of the amplitude of the resonant or nonresonant contribution $i$;
\item $c_i$ and $\theta_i$ are the real parameters to be determined ($-\pi \leq \theta_i \leq \pi$);
\item $\epsilon(m_{13}^2,m_{23}^2)$ is the reconstruction efficiency defined for all points in the Dalitz plot;
\item $Q(m_{13}^2,m_{23}^2)$ is the distribution of \qqbar\ continuum background;
\item $B(m_{13}^2,m_{23}^2)$ is the distribution of \BB\ background; and
\item $f_{\qqbar}$ and $f_{\BB}$ are the fractions of \qqbar\ continuum and \BB\ background events, respectively.
      They are determined from the \mes\ fit and MC, respectively, and are fixed in the amplitude fit.
\end{itemize}

The first term on the right-hand-side in Eq.~\ref{LikeEqn} 
corresponds to the signal probability density
function (PDF) multiplied by the signal fraction
$(1 - f_{\qqbar} - f_{\BB})$. Since we can always apply a common
factor to both the numerator and demoninator of the signal PDF, this analysis
will only be sensitive to relative phases and magnitudes, and
hence it is possible to fix the magnitude and phase of one component 
(\rhoI\ for \PPP\ and \KstarI\ for \KPP).

As the choice of normalisation, phase convention and amplitude
formalism may not always be the same for different experiments, 
fit fractions are presented instead of amplitude magnitudes to allow a more
meaningful comparison of results. The fit fraction is defined as the
integral of a single decay amplitude squared divided by the coherent matrix
element squared for the complete Dalitz plot as shown in Eq.~\ref{eqn:fitfraction}.

\begin{equation}
{\mbox{Fit Fraction}} = 
\frac{\int|c_i e^{i\theta_i}F_i(m_{13}^2, m_{23}^2)|^2dm_{13}^2dm_{23}^2}{\int |\sum_i c_i e^{i\theta_i} F_i(m_{13}^2, m_{23}^2)|^2 dm_{13}^2dm_{23}^2}.
\label{eqn:fitfraction}
\end{equation}

Note that the sum of these fit fractions
is not necessarily unity due to the potential presence of net constructive
or destructive interference.

%% file: results_pipipi.tex
\subsection{\boldmath \BtoPPP\ Results}
\label{sec:pipipi}

The nominal fit is performed with the resonances \rhoI, \rhoII, \fz,
\fzII, and a uniform nonresonant (NR) contribution.
The masses and widths of the resonances are fixed to their world average values~\cite{pdg2004}.  
In this fit the \rhoI\ is fixed to have a magnitude of 1 and its phase is set to 0, since this is the
dominant contribution to the Dalitz plot, and this choice reduces
the statistical uncertainties for the other fitted components.
We model all the resonances using relativistic Breit--Wigner lineshapes 
with Blatt--Weisskopf barrier factors~\cite{blatt} 
except for the \fz, which is modeled with a Flatt\'e lineshape~\cite{flatte} 
(to account for its coupled-channel behaviour due to the fact that it can decay to $\pipi$ or $\Kp\Km$).
The nonresonant component is assumed to be uniform in phase space.  
The preliminary results of the nominal fit to the 2407 events in the signal box can be seen in 
Table~\ref{tab:pipipi_nominal}, along with the total branching fraction (BF) 
and the average efficiency across the Dalitz plot
weighted by the fitted signal distribution.

\input Tables/nominal3piTable_all

\begin{figure}[!ht]
\begin{center}
\includegraphics[width=1.0\textwidth]{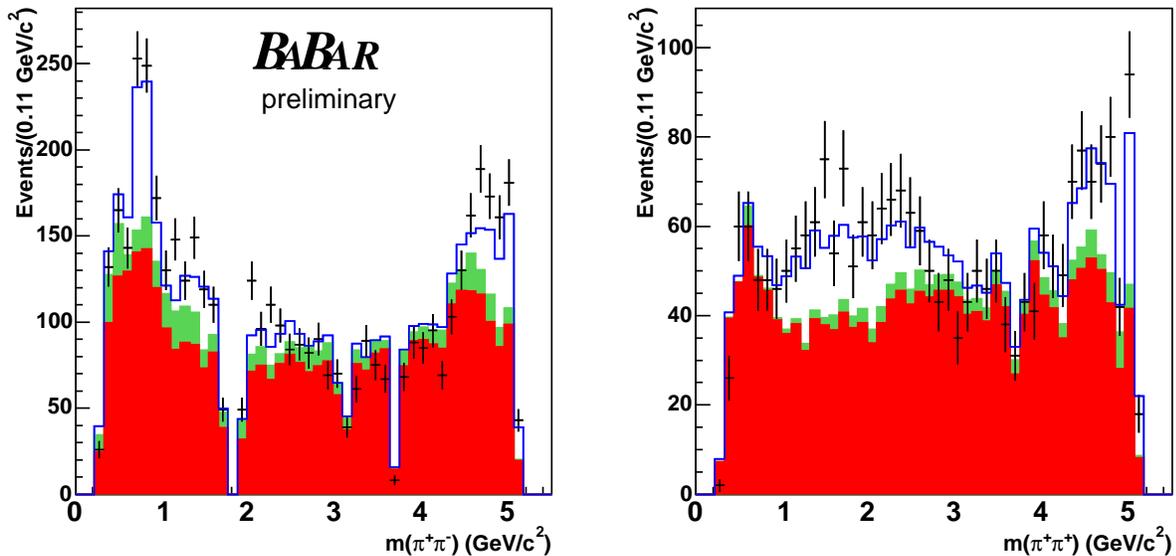}
\caption{Projection plots of the fit result for \BtoPPP\ 
onto the mass variables $m_{\pi^+ \pi^-}$ and $m_{\pi^{\pm} \pi^{\pm}}$.
The data are the black points with error bars, the lower solid red histogram is the
\qqbar\ component, the middle solid green histogram is the
\BB\ background contribution, while the upper blue histogram shows
the total fit result. All errors shown are statistical only.
The large dips in the spectra correspond to the vetoes in Table~\ref{tab:veto}.}
\label{fig:Fit3piData}
\end{center}
\end{figure}

Figure~\ref{fig:Fit3piData} shows the mass projection plots for the nominal fit.
The four resonant contributions plus the
single uniform phase-space nonresonant model are able to adequately
describe the data within the statistical uncertainties. 

Further fits are performed to the data by removing one two-body component
at a time from the nominal model; the results are shown in
Table~\ref{tab:omit3pitab}. Removing the $\rhoI$, $\rhoII$ or
$\fzII$ components give significantly worse fit results. The
omission of the $\fz$ and nonresonant amplitudes
give values of the remaining fitted components that are
close to their nominal values.

Recent experimental results from $e^+ e^-$ collisions at BES~\cite{BESsigma}
show evidence of a low-mass $\pi^+ \pi^-$ pole in data for
$J/\psi \rightarrow \omega \pi^+ \pi^-$, known as the $\sigma$. 
Analysis of data from the E791 experiment for 
$D^+ \rightarrow \pi^+ \pi^- \pi^+$~\cite{E791} show similar results.
Also a large concentration of events
in the $I=0$ S-wave $\pi \pi$ channel has been seen in the $m_{\pi \pi}$ region around $500 - 600$ MeV in $pp$ collisions~\cite{Alde}. This pole is
predicted from models based on chiral perturbation theory~\cite{Colangelo}, in
which the resonance parameters are
$M -i\Gamma/2 = \left[ (470 \pm 30) - i(295 \pm 20) \right] $ MeV.

Consequently the $\sigma$ resonance is predicted in 
$B^{\pm} \rightarrow \pi^+ \pi^- \pi^{\pm}$ decays. For this Dalitz-plot
analysis the $\sigma$ resonance is modeled using the parameterisation
suggested by Bugg~\cite{bugg}.
The result of the Dalitz-plot fit is shown in Table~\ref{tab:add3pitab}. It 
can be seen that the inclusion of the $\sigma$ slightly changes
the contributions of the other resonant components and also affects
the phase of the nonresonant amplitude, but all results are consistent with those
from the nominal fit, within the statistical uncertainties.

Table~\ref{tab:add3pitab} also shows the preliminary results of the fit to the data when
we add the $\fzIII$ and $\chiczero$ resonances to the model. Both contributions
are found to be negligible.

\input Tables/omit3piTable_all

\input Tables/add3piTable_all

\clearpage

%% file: Tables/nominal3piTable_all.tex
\begin{table}[!h]
\caption{Results of the nominal fit for the \BtoPPP\ mode. The first errors
are statistical, while the second errors are systematic and are described in
Section~\ref{sec:Systematics}.}
\label{tab:pipipi_nominal}
\begin{center}
\begin{tabular}{|l|c|}
\hline
Average Efficiency (\%) & $13.0 \pm 0.9 \pm 0.6$ \\
\hline
Total BF ($\times10^{-6}$) & $16.2 \pm 2.1 \pm 1.3$\\
\hline
\hline
Component  & Fit Result\\
\hline
\hline
\rhoI\  Fit Fraction (\%) & $58.2 \pm 2.9 \pm 6.0$ \\
\rhoI\ Phase& 0.0 (Fixed) \\
\hline
\rhoII\  Fit Fraction (\%) & $13.6 \pm 2.8 \pm 2.0$ \\
\rhoII\ Phase&     $+0.59 \pm 0.39 \pm 0.17$ \\
\hline
\fz\  Fit Fraction (\%) &    $2.0 \pm 1.3 \pm 2.8$ \\
\fz\ Phase&         $+2.45 \pm 0.61 \pm 0.19$ \\
\hline
\fzII\  Fit Fraction (\%) &       $14.3 \pm 2.0 \pm 1.8$ \\
\fzII\ Phase&           $-2.69 \pm 0.33 \pm 0.17$ \\
\hline
NR  Fit Fraction (\%) &        $4.2 \pm 2.0 \pm 1.4$ \\
NR Phase&            $+0.61 \pm 0.56 \pm 0.19$ \\
\hline
\end{tabular}
\end{center}
\end{table}

%% file: Tables/omit3piTable_all.tex
\begin{table}[!h]
\caption{Results of the fit to the Dalitz plot for \BtoPPP\ candidates, 
for the nominal fit and for fits performed with a different component omitted in turn from the nominal fit.
All errors shown are statistical only.}
\label{tab:omit3pitab}
\begin{center}
\resizebox{\textwidth}{!}{
\begin{tabular}{|l|c|c|c|c|c|c|}
\hline
  & Nominal & No \rhoI & No \rhoII & No \fz & No \fzII & No NR \\
\hline
$(-\rm{ln}{\cal{L}}) - (-\rm{ln}{\cal{L}}(\rm{nominal}))$ & ---   & 94.3  & 8.0   & 1.7   & 18.9  & 2.8 \\

\hline
$\rhoI$ Fit Fraction (\%)  & $58.2 \pm 2.9$ & --- & $71.2 \pm 3.0$ & $58.4 \pm 2.6$ & $66.4 \pm 3.8$ & $58.9 \pm 2.5$ \\

\hline
$\rhoII$ Fit Fraction (\%) & $13.6 \pm 2.8$ & $46.2 \pm 3.5$ & --- & $14.8 \pm 3.0$ & $16.1 \pm 3.6$ & $16.9 \pm 2.9$\\
$\rhoII$ Phase       & $+0.59 \pm 0.39$& $+0.59$ (Fixed) & --- & $+0.58 \pm 0.39$ & $+0.30 \pm 0.41$ & $+0.46 \pm 0.36$ \\

\hline
$\fz$ Fit Fraction (\%) & $2.0 \pm 1.3$ & $21.9 \pm 2.9$ & $4.5 \pm 1.3$ & --- & $1.6 \pm 1.8$ & $1.5 \pm 1.0$ \\
$\fz$ Phase & $+2.45 \pm 0.61$  & $+2.58 \pm 0.29$  & $+2.92 \pm 0.39$   & --- & $+2.27 \pm 0.83$ & $+2.78 \pm 0.59$ \\

\hline
$\fzII$ Fit Fraction (\%) & $14.3 \pm 2.0$ & $23.2 \pm 2.6$ & $16.7 \pm 1.4$ & $14.1 \pm 2.1$ & --- & $14.8 \pm 2.0$\\
$\fzII$ Phase & $-2.69 \pm 0.33$ & $-2.95 \pm 0.24$ & $-2.09 \pm 0.22$ & $-2.75 \pm 0.35$ & --- & $-2.84 \pm 0.33$ \\

\hline
NR Fit Fraction (\%) & $4.2 \pm 2.0$ & $34.2 \pm 3.7$ & $11.8 \pm 2.5$ & $3.1 \pm 1.7$ & $6.2 \pm 2.7$ & --- \\
NR Phase & $+0.61 \pm 0.56$ & $+1.84 \pm 0.23$  & $+0.82 \pm 0.39$ & $+0.27 \pm 0.58$ & $+0.07 \pm 0.56$ & --- \\

\hline
\end{tabular}
}
\end{center}
\end{table}

%% file: Tables/add3piTable_all.tex
\begin{table}[!h]
\caption{Results of the fit to the Dalitz plot for \BtoPPP\ candidates, 
for the nominal fit and for fits performed with a different component added in turn to the nominal fit.
All errors shown are statistical only.}
\label{tab:add3pitab}
\begin{center}
\begin{tabular}{|l|c|c|c|c|}
\hline
  & Nominal & With $\sigma$ & With $\fzIII$ & With $\chiczero$ \\
\hline
$(-\rm{ln}{\cal{L}}) - (-\rm{ln}{\cal{L}}(\rm{nominal}))$ & ---  & -3.9 & -0.1 & 0.0\\

\hline
$\rhoI$ Fit Fraction (\%)  & $58.2 \pm 2.9$ & $52.8 \pm 6.4$ & $58.4 \pm 3.7$ & $58.1 \pm 5.2$ \\

\hline
$\rhoII$ Fit Fraction (\%) & $13.6 \pm 2.8$ & $10.6 \pm 3.2$ & $13.6 \pm 2.8$ & $13.6 \pm 3.0$\\
$\rhoII$ Phase       & $+0.59 \pm 0.39$ & $+0.88 \pm 0.46$ & $+0.58 \pm 0.38$ & $+0.60 \pm 0.39$ \\

\hline
$\fz$ Fit Fraction (\%) & $2.0 \pm 1.3$ & $5.9 \pm 1.6$ & $1.7 \pm 1.8$ & $1.9 \pm 1.3$ \\
$\fz$ Phase & $+2.45 \pm 0.61$  & $+2.19 \pm 0.37$  & $+2.32 \pm 0.79$ & $+2.45 \pm 0.61$ \\

\hline
$\fzII$ Fit Fraction (\%) & $14.3 \pm 2.0$ & $12.4 \pm 2.5$ & $14.2 \pm 2.1$ & $14.2 \pm 2.3$ \\
$\fzII$ Phase & $-2.69 \pm 0.33$ & $-2.44 \pm 0.36$ & $-2.69 \pm 0.33$ & $-2.69 \pm 0.33$ \\

\hline
NR Fit Fraction (\%) & $4.2 \pm 2.0$ & $1.0 \pm 2.6$ & $3.8 \pm 2.7$ & $4.1 \pm 2.0$ \\
NR Phase & $+0.61 \pm 0.56$ & $+2.99 \pm 1.06$  & $+0.69 \pm 0.71$ & $+0.61 \pm 0.56$ \\

\hline

$\sigma$ Fit Fraction (\%) & --- & $25.7 \pm 9.1$ & --- & --- \\
$\sigma$ Phase & --- & $-1.62 \pm 0.21$ & --- & --- \\

\hline

$\fzIII$ Fit Fraction (\%) & --- & --- & $0.0 \pm 2.1$ & \\
$\fzIII$ Phase & --- & --- & $-1.7 \pm 2.2$ & \\

\hline

$\chiczero$ Fit Fraction (\%) & --- & --- & --- & $0.2 \pm 7.5$ \\
$\chiczero$ Phase & --- & --- & --- & $-0.7 \pm 4.7$ \\

\hline

\end{tabular}
\end{center}
\end{table}

%% file: results_kpipi.tex
\subsection{\boldmath \BtoKPP\ Results}
\label{sec:kpipi}

The expected significant contributions to the \KPP\ Dalitz plot can be identified from previous studies of this final state. 
Our model includes
the following resonances: \KstarI, \KstarII, \rhoI, \fz, \chiczero\ and a nonresonant amplitude. 
The masses and widths of the resonances are fixed to their world average values~\cite{pdg2004}.  
In the fits we use \KstarI\ as the reference component and hence its
magnitude and phase are fixed to 1 and 0, respectively.  
We use the same lineshapes described in Sec.~\ref{sec:pipipi} to model the resonance dynamics, except for \KstarII, where we use the LASS amplitude model.
The dynamics of the $K\pi$ S-wave are not very well established 
and form an area of some disagreement within the community.  
Some favor the existence of the $\kappa$ pole~\cite{bugg}, 
whilst others strongly oppose it.  
All, however, agree that there is strong evidence of resonant behavior 
around 1430 \mev, the \KstarII.  
The LASS experiment studied $K\pi$ scattering and as part of this study 
produced a description of the S-wave that consists of a resonant part, 
the \KstarII, and an effective-range term~\cite{LASS,billnote}.  
This amplitude is only measured up to around 2 \gevcc\ in $K\pi$ mass, 
and so we curtail the effective-range term at the lower edge of the \Dz\ veto.  
Since the LASS amplitude contains both a resonant and nonresonant part, 
the results for \KstarII\ are not purely due to this resonance, 
but to the $K\pi$ S-wave as a whole.
This model, with five two-body components plus a uniform nonresonant component, with the LASS amplitude for the \KstarII\
will be referred to as the ``nominal'' model.

\begin{figure}[ht!]
\begin{center}
\includegraphics[width=\textwidth]{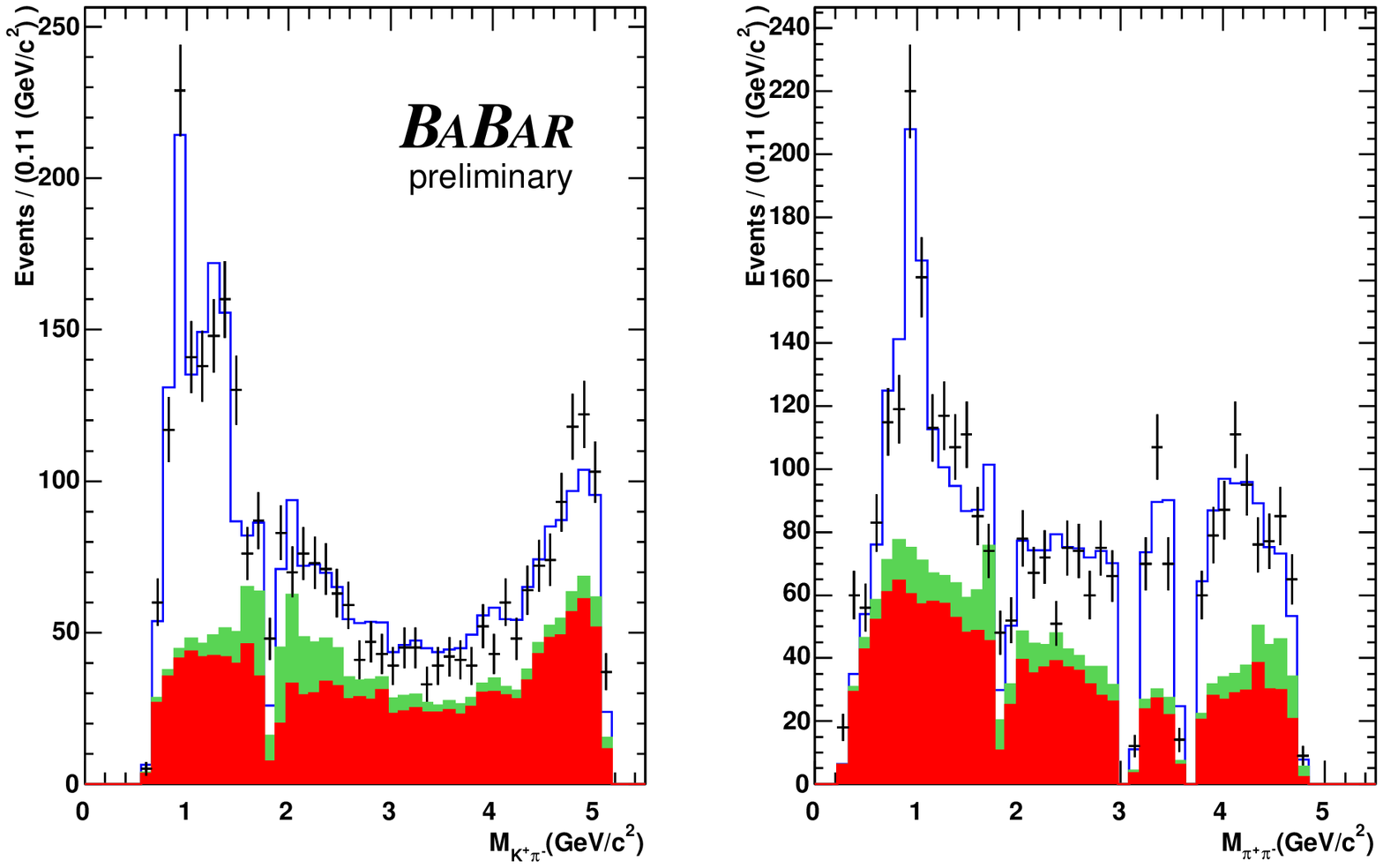}
\caption[Mass pair projection plots for the nominal \BtoKPP\ fit.]
        {Mass pair projection plots for the nominal \BtoKPP\ fit.
         The left plot shows the $\Kpm\pimp$ mass spectrum and the right plot shows the $\pipm\pimp$ mass spectrum.
         The data are the black points with error bars,
         the lower solid red histogram is the \qqbar\ component, 
         the middle solid green histogram is the \BB\ background contribution, 
         while the upper blue histogram shows the total fit result. 
         All errors shown are statistical only.
         The large dips in the spectra correspond to the vetoes in Table~\ref{tab:veto}.}
\label{fig:invmass_data}
\end{center}
\end{figure}

The nominal fit shows very good agreement with the data; 
a comparison can be seen in Figure~\ref{fig:invmass_data}.
The preliminary results of the nominal fit to the 3174 events in the signal box can be seen in Table~\ref{tab:kpipi_nominal} 
along with the total branching fraction (BF) and the average efficiency across the Dalitz plot
weighted by the fitted signal distribution.
We have also used a relativistic Breit-Wigner
or a Flatt\'e lineshape for the \KstarII\ resonance 
with and without the addition of a $\kappa$ resonance, as suggested in Ref~\cite{bugg}, 
to model the $K\pi$ S-wave
but found the fits to be poor representations of the data. 

\input nominalTable

We refit the data removing each component in turn and find, in each case, 
that the fit worsens considerably.
In some cases the parameters of the other components can vary dramatically.
Table~\ref{tab:omittab} details the results of these omission tests.
We also test for the possibility that there are further resonances in
the Dalitz plot by repeating the fit using the nominal model but adding
an additional component.
The possible additional states we test for are the \fzII, \fzIII\ 
and \rhoII\ resonances in the $\pi\pi$ mass spectrum and the 
\KstarIII, \KstarIV\ and $\kappa$ resonances in the $K\pi$ mass spectrum.  
The results of these fits are shown in Table~\ref{tab:addtab}.  
In all cases the \NLL\ is slightly better than in the original fit.
These checks are performed only on the \Bp\ sample.

\input omitTable

\input addTable

\clearpage

%% file: nominalTable.tex
\begin{table}[!h]
\caption{Results of the nominal fit for the \BtoKPP\ mode. 
The first errors are statistical, while the second errors are systematic,
and are described in Section~\ref{sec:Systematics}.}
\label{tab:kpipi_nominal}
\begin{center}
\begin{tabular}{|l|c|}
\hline
Average Efficiency (\%) & $12.8\pm0.8\pm0.5$ \\
\hline
Total BF ($\times10^{-6}$) & $61.4\pm2.4\pm4.5$\\
\hline
\hline
Component  & Fit Result\\
\hline
\hline
\KstarI\  Fit Fraction (\%) & $11.4\pm 2.0\pm1.5$ \\
\KstarI\ Phase& 0.0 (Fixed) \\
\hline
\KstarII\  Fit Fraction (\%) &$52.6\pm2.3\pm4.0$ \\
\KstarII\ Phase&     $+2.92 \pm 0.11 \pm 0.10$ \\
\hline
\rhoI\  Fit Fraction (\%) &    $8.5\pm1.9\pm1.1$ \\
\rhoI\ Phase&         $+0.85 \pm 0.38 \pm 0.35$\\
\hline
\fz\  Fit Fraction (\%) &       $15.0\pm2.4\pm1.3$ \\
\fz\ Phase&           $-0.55 \pm 0.32 \pm 0.41$ \\
\hline
\chiczero\  Fit Fraction (\%) & $1.45\pm0.27\pm0.23$ \\
\chiczero\ Phase&     $+0.15 \pm 0.33 \pm 0.22$ \\
\hline
NR  Fit Fraction (\%) &        $7.9\pm0.9\pm2.3$ \\
NR Phase&            $+0.50 \pm 0.24 \pm 0.24$\\
\hline
\end{tabular}
\end{center}
\end{table}

%% file: omitTable.tex
\begin{sidewaystable}[!h]
\caption{Results of the fit to the Dalitz plot for \BtoKpppos\ candidates, 
for the nominal fit and for fits performed with a different component omitted in turn from the nominal fit.
All errors shown are statistical only.}
\label{tab:omittab}
\begin{center}
\resizebox{\textheight}{!}{
\begin{tabular}{|l|c|c|c|c|c|c|c|}
\hline
  & Nominal & No \KstarI & No \KstarII & No \rhoI & No \fz & No \chiczero & No NR \\
\hline
(\NLL) $-$ (\NLL(nominal))      & ---      & 55.4     & 156.9   & 25.0     & 79.9   & 17.0   & 21.5   \\

\hline
$\KstarI$ Fit Fraction (\%)  & $9.5\pm1.8$ & --- & $19.9\pm1.1$ & $9.8\pm1.8$ & $10.3\pm4.0$ & $9.6\pm1.7$ & $9.2\pm1.5$ \\

\hline
$\KstarII$ Fit Fraction (\%) & $54.3\pm3.1$ & $64.5\pm9.6$ & --- & $59.4\pm3.8$ & $58.8\pm3.9$ & $54.8\pm3.1$ & $54.2\pm2.9$ \\
$\KstarII$ Phase       & $+2.98\pm0.16$& $+3.03\pm0.34$ & --- & $+2.96\pm0.16$ & $+3.03\pm0.15$ & $+3.03\pm0.16$ & $+2.89\pm0.15$ \\

\hline
$\rhoI$ Fit Fraction (\%) & $8.6\pm1.9$ & $8.6\pm1.1$ & $13.1\pm1.5$ & --- & $14.4\pm5.0$ & $8.2\pm1.7$ & $11.0\pm2.5$\\
$\rhoI$ Phase & $+1.48\pm0.45$  & $+1.73\pm0.28$  &$+1.68\pm0.30$   & --- & $+1.55\pm0.57$ & $+1.67\pm0.43$ & $+0.45\pm0.45$  \\

\hline
$\fz$ Fit Fraction (\%) & $13.4\pm2.4$ & $13.3\pm2.0$ & $32.3\pm2.3$ & $16.8\pm3.0$ & --- & $12.7\pm2.2$ & $20.8\pm3.2$\\
$\fz$ Phase & $-0.18\pm0.40$ & $-0.18$ (Fixed) &$+0.18\pm0.21$ & $+0.06\pm0.40$ & --- & $+0.05\pm0.39$ & $-1.45\pm0.37$ \\

\hline
$\chiczero$ Fit Fraction (\%) & $1.57\pm0.41$ & $1.57\pm0.41$ & $1.46\pm0.24$ & $1.57\pm0.41$ & $1.38\pm0.39$ & --- & $2.27\pm0.56$ \\
$\chiczero$ Phase & $+0.75\pm0.47$ & $+1.06\pm0.46$   & $-2.45\pm0.34$ & $+0.67\pm0.47$     & $+0.84\pm0.48$ & --- & $+1.08\pm0.47$ \\

\hline
NR Fit Fraction (\%) & $9.5\pm1.3$ & $7.35\pm0.97$ & $43.2\pm2.6$ & $13.7\pm1.8$ & $14.1\pm1.8$ & $12.2\pm1.5$ & --- \\
NR Phase & $+0.71\pm0.32$ & $+1.16\pm0.27$  &$-1.62\pm0.17$ & $+0.66\pm0.32$ & $+0.99\pm0.32$ & $+0.89\pm0.30$ & --- \\   

\hline
\end{tabular}
}
\end{center}
\end{sidewaystable}

%% file: addTable.tex
\begin{sidewaystable}[!h]
\caption{Results of the fit to the Dalitz plot for \BtoKpppos\ candidates, 
for the nominal fit and for fits performed with a different component added in turn to the nominal fit.
All errors shown are statistical only.}
\label{tab:addtab}
\begin{center}
\resizebox{\textheight}{!}{
\begin{tabular}{|l|c|c|c|c|c|c|c|}
\hline
  & Nominal & With \fzII & With \fzIII & With \rhoII & With \KstarIV & With \KstarIII & With $\kappa$\\
\hline
(\NLL) $-$ (\NLL(nominal))      & ---      &  $-2.1$   & $-2.1$   & $-1.9$     & $-0.7$   & $-3.7$ & $-3.7$\\

\hline
$\KstarI$ Fit Fraction (\%)  & $9.5\pm1.8$ & $9.0\pm2.4$ & $9.5\pm2.2$ & $9.2\pm2.7$ & $9.2\pm1.9$ & $9.8\pm2.1$ & $9.4\pm2.1$\\
\hline
$\KstarII$ Fit Fraction (\%) & $54.3\pm3.1$ & $53.5\pm3.7$ & $53.7\pm3.6$ & $55.5\pm4.6$ & $53.3\pm3.4$ & $50.7\pm3.1$ & $54.4\pm3.6$\\
$\KstarII$ Phase       & $+2.98\pm0.16$& $+3.02\pm0.16$ & $+3.00\pm0.16$ & $+2.92\pm0.17$ & $+2.97\pm0.17$ & $+3.00\pm0.16$&$2.97\pm0.16$\\
\hline
$\rhoI$ Fit Fraction (\%) & $8.6\pm1.9$ & $7.7\pm1.8$ & $7.7\pm1.7$ & $9.0\pm1.3$ & $8.5\pm2.0$ & $8.0\pm1.7$ & $7.1\pm1.6$\\
$\rhoI$ Phase & $+1.48\pm0.45$  & $+1.47\pm0.45$  &$+1.66\pm0.44$   & $+1.24\pm0.55$ & $+1.50\pm0.46$ & $+1.87\pm0.43$& $+1.20\pm0.44$\\
\hline
$\fz$ Fit Fraction (\%) & $13.4\pm2.4$ & $13.8\pm3.5$ & $11.0\pm2.5$ & $14.3\pm4.1$ & $13.6\pm2.6$ & $13.3\pm2.7$ & $12.8\pm2.6$\\
$\fz$ Phase & $-0.18\pm0.40$ & $-0.14\pm0.48$   &$+0.22\pm0.44$ & $+0.18\pm0.51$ & $-0.20\pm0.42$& $+0.15\pm0.44$& $-0.62\pm0.44$ \\
\hline
$\chiczero$ Fit Fraction (\%) & $1.57\pm0.41$ & $1.53\pm0.41$ & $1.57\pm0.42$ & $1.61\pm0.46$ & $1.58\pm0.43$ &$1.55\pm0.43$ & $1.61\pm0.42$\\
$\chiczero$ Phase & $+0.75\pm0.47$    & $+0.77\pm0.47$   & $+0.77\pm0.47$ & $+0.54\pm0.49$     & $+0.80\pm0.48$& $+0.87\pm0.48$& $+0.72\pm0.47$ \\
\hline
NR Fit Fraction (\%) & $9.5\pm1.3$ & $9.5\pm1.3$ & $11.5\pm1.6$ & $11.0\pm2.0$ & $9.0\pm1.3$ & $8.6\pm1.5$ & $9.9\pm1.3$\\
NR Phase & $+0.71\pm0.32$ & $+0.75\pm0.31$  &$+0.74\pm0.32$ & $+0.53\pm0.38$ & $+0.73\pm0.34$& $+1.07\pm0.38$& $+0.74\pm0.31$ \\   
\hline
$\fzII$ Fit Fraction (\%) & --- & $2.1\pm1.9$ & --- & --- & --- &--- & ---\\
$\fzII$ Phase & ---  & $-0.09\pm0.49$   & --- & --- & --- & ---& ---\\
\hline
$\fzIII$ Fit Fraction (\%) & --- & --- & $1.15\pm0.74$ & --- & --- &--- & ---\\
$\fzIII$ Phase & --- & ---   &$+0.38\pm0.64$ & --- & --- &---& ---\\
\hline
$\rhoII$ Fit Fraction (\%) & --- & --- & --- & $6.5\pm2.5$ & --- &---& ---\\
$\rhoII$ Phase & --- & ---   & --- & $-0.92\pm0.56$ & --- &---& ---\\
\hline
$\KstarIV$ Fit Fraction (\%) & --- & --- & --- & --- & $0.71\pm1.24$ &--- & ---\\
$\KstarIV$ Phase & --- & ---   & --- & --- & $+2.35\pm0.85$ &---& ---\\
\hline
$\KstarIII$ Fit Fraction (\%) & --- & --- & --- & --- & --- & $3.95\pm0.54$ & ---\\
$\KstarIII$ Phase & --- & ---   & --- & --- & --- & $-0.35\pm0.25$& ---\\
\hline
$\kappa$ Fit Fraction (\%) & --- & --- & --- & --- & --- & --- & $3.03\pm0.89$\\
$\kappa$ Phase & --- & ---   & --- & --- & --- & ---& $+2.90\pm0.42$\\
\hline
\end{tabular}
}
\end{center}
\end{sidewaystable}

%% file: systematics.tex
\section{Systematic Studies}
\label{sec:Systematics}

The charged particle tracking and particle identification uncertainties are 
2.4\% and 3.0\%, respectively.
There are also global systematic errors in the efficiencies due to the
criteria applied to the event-shape variables (1.0\%) and to $\Delta E$ and $m_{ES}$ (1.0\%). 
We also take into account the statistical uncertainty on the efficiency 
(due to the weighting by the amplitude model)
for the individual and total branching fraction results
(7\% for \BtoPPP, 6\% for \BtoKPP).
The uncertainty in the number of \BB\ events is evaluated to be 1.1\%.

The systematic error on the efficiency variation across the Dalitz 
plot is calculated by performing a series of fits to the data where 
we vary the contents of each bin in the efficiency histogram according to
binomial errors. This introduces an absolute uncertainty
of $0.02$ to $0.09$ for the phases, and a fractional uncertainty
of $0.4$\% to $4.9\%$ for the fit fractions. However, 
for the average efficiency, and hence for the total branching
fraction, this is a very small effect, evaluated as 0.1\%.

The systematic uncertainty introduced by the \BB\ background and \qqbar\ 
background has two components, 
each of which can potentially affect the fitted magnitudes and phases differently.
The first component arises from the uncertainty in the overall normalisation of these backgrounds,
whilst the second component arises from the uncertainty on the shapes of the background distributions in the Dalitz plot.
The uncertainties on the fit fractions and phases due to the normalisation uncertainty 
are estimated by varying the measured background fractions in the signal box by their statistical errors. 
The maximum associated absolute uncertainty for the phase is $0.85$ ($0.05$)
for $\PPP$ ($\KPP$) due to the \qqbar\ background normalisation uncertainty  
and $0.03$ due to the \BB\ background normalisation uncertainty. These uncertainties are added in quadrature.
The fit fractions are affected in a less uniform manner, 
with relative uncertainties in the range 
$0.7$\% to $15.4$\% ($0.2$\% to $4.1$\%) for $\PPP$ ($\KPP$).
The uncertainties on the fit fractions and phases due to the Dalitz-plot background distribution uncertainty 
is estimated in the same way as the efficiency variation, 
namely varying the contents of the histogram bins in accordance with their Poisson errors.
To be conservative, each phase for \PPP\ has been given an associated uncertainty 
of 0.1 due to the \qqbar\ background distribution uncertainty  
and 0.1 due to the \BB\ background distribution uncertainty, 
which are then added in quadrature.
For \KPP\ the variations between modes are larger and each mode is therefore treated individually.
The range of uncertainties in the phases is $0.06$ to $0.36$ due to \qqbar\ background and
$0.04$ to $0.18$ due to \BB\ background, which are again added in quadrature.
The fit fractions are affected in a less uniform manner, 
with the relative uncertainties ranging from 
$1$\% to $8$\% ($1$\% to $25$\%) for \PPP\ (\KPP).

Possible biases due to the fitting procedure are investigated using
extensive Monte Carlo simulations. We assign an absolute systematic error up to 3\%
for the fit fractions and 0.1 for the phases.

%% file: summary.tex
\section{Summary}
\label{sec:Summary}

Tables~\ref{tab:pipipi_nominal} and~\ref{tab:kpipi_nominal} show the preliminary results 
from the nominal fits together with their statistical and systematic errors.

For the decay \BtoPPP\ the nominal fit to the Dalitz plot is performed
with the resonances \rhoI, \rhoII, \fz, \fzII\ and a uniform nonresonant
contribution. The total branching fraction is consistent with the previously 
measured value~\cite{Babar1, pdg2004}. We can estimate the branching
fraction of $\modeIII$ by multiplying its fit fraction by the total branching
fraction for \BtoPPP. We find this to be $(9.4 \pm 1.3 \pm 1.0) \times 10^{-6}$,
which is consistent with the previously measured value~\cite{rhopi, pdg2004}.
The removal of any of the fit components
gives worse likelihood values, especially for the $\rhoI$, $\rhoII$ and $\fzII$
resonances. It is found that the \fzIII\ and \chiczero\ resonances have no contribution
to this Dalitz plot, while there is some evidence for a contribution from the $\sigma$.

For the decay \BtoKPP\ the nominal fit to the Dalitz plot is performed
with the resonances \KstarI, \KstarII, \rhoI, \fz, \chiczero, 
a uniform nonresonant component and the LASS parameterization of the
scalar component of the $K\pi$ spectrum.
The total branching fraction result is consistent with the previously
measured value~\cite{Babar1, pdg2004}.
We can estimate the branching fraction of $\Bpm \to \KstarI \pipm, \KstarI \to \Kpm \pimp$
by multiplying its fit fraction by the total branching fraction for \BtoKPP. 
This yields a value of $(7.0 \pm 1.3 \pm 0.9) \times 10^{-6}$,
which is smaller than that reported by earlier
analyses that do not fit over the full Dalitz region but is consistent
with the Dalitz analysis reported by Belle~\cite{belleDalitz}. 
The resonance behavior around $1400$\mev\ is successfully modeled 
with the LASS parameterization of the $K\pi$ S-wave. 
The removal of any of the fit components results in a significant worsening of 
the fit likelihood and a large change in one or more of the remaining amplitudes. 
Addition of further resonances (\fzII, \fzIII, \rhoII, \KstarIII, \KstarIV\ or $\kappa$) 
does not cause a significant change in the fit likelihood, and in most cases
the extra component is measured to be compatible with having zero fit fraction.  
In addition, their presence in the fit does not significantly alter any of the other parameters 
and consequently has no significant effect on the fit fractions of the original components.
The fit is stable with respect to different parameterizations of the Flatt\'e line shape of the \fz\ .

Future iterations of the analysis will use the information from the separate \Bp\ and \Bm\ samples 
to make measurements of the charge asymmetries of the various intermediate decay modes, 
potentially allowing observation of direct \CP\ violation.

%% file: pubboard/acknowledgements.tex
We are grateful for the 
extraordinary contributions of our \pep2\ colleagues in
achieving the excellent luminosity and machine conditions
that have made this work possible.
The success of this project also relies critically on the 
expertise and dedication of the computing organizations that 
support \babar.
The collaborating institutions wish to thank 
SLAC for its support and the kind hospitality extended to them. 
This work is supported by the
US Department of Energy
and National Science Foundation, the
Natural Sciences and Engineering Research Council (Canada),
Institute of High Energy Physics (China), the
Commissariat \`a l'Energie Atomique and
Institut National de Physique Nucl\'eaire et de Physique des Particules
(France), the
Bundesministerium f\"ur Bildung und Forschung and
Deutsche Forschungsgemeinschaft
(Germany), the
Istituto Nazionale di Fisica Nucleare (Italy),
the Foundation for Fundamental Research on Matter (The Netherlands),
the Research Council of Norway, the
Ministry of Science and Technology of the Russian Federation, and the
Particle Physics and Astronomy Research Council (United Kingdom). 
Individuals have received support from 
CONACyT (Mexico),
the A. P. Sloan Foundation, 
the Research Corporation,
and the Alexander von Humboldt Foundation.

%% file: Dalitz-ICHEP-2004-Paper.bbl
\begin{thebibliography}{99}


\bibitem{Belle}
Belle Collaboration, K. ~Abe {\it et al.},
\jprd{65}, 092005 (2002)

\bibitem{Babar1}
\babar\ Collaboration, B.~Aubert {\it et al.},
presented at FPCP 2002, hep-ex/0206004; \\
\babar\ Collaboration, B. Aubert {\it et al.},
\jprl{91}, 051810 (2003)

\bibitem{Belle2}
Belle Collaboration, K.~Abe {\it et al.}, 
\jprd{69}, 012001 (2004)

\bibitem{belleDalitz}
Belle Collaboration, K.~Abe {\it et al.},
presented at HEP 2003, BELLE-CONF-0338

\bibitem{babardet}
\babar\ Collaboration, B.\ Aubert {\em et al.}, ``The \babar\ Detector,''
\nima{479}, 1 (2002)

\bibitem{pep}
PEP-II Conceptual Design Report, SLAC-R-418 (1993)




\bibitem{fisher}
R.A.~Fisher,
Annals Eugen. {\bf 7}, 179 (1936);\\
G.~Cowan,
\emph{Statistical Data Analysis}, (Oxford University Press, 1998), p51

\bibitem{CLEO}
CLEO Collaboration, D.\ M.\ Asner, {\it et al.},
\jprd{53}, 1039 (1996)


\bibitem{argus}
ARGUS Collaboration, H.~Albrecht, {\it et al.},
\zp{C48}, 543 (1990)



\bibitem{pdg2004}
Particle Data Group, S.\ Eidelman {\it et al.},
\plb{592}, 1 (2004)

\bibitem{blatt}
J.~Blatt and V.~Weisskopf, 
``Theoretical Nuclear Physics'', 
John Wiley and Sons, New York, 1956.

\bibitem{flatte} 
S.M.~Flatt\'e, 
\plb{63}, 224 (1976)

\bibitem{BESsigma}
BES Collaboration, 
``The $\sigma$ pole in $\jpsi \rightarrow \omega \pipi$'', submitted to \plb, hep-ex/0406038

\bibitem{Alde}
D.~Alde {\it et al.}, 
\plb{397}, 350 (1997)

\bibitem{Colangelo}
G.~Colangelo {\it et al.}, 
\npb{603}, 125 (2001)

\bibitem{E791}
E791 Collaboration, E.M.~Aitala {\it et al.}, 
\jprl{86}, 770 (2001)

\bibitem{bugg}
D.V.~Bugg,
\plb{572}, 1 (2003)

\bibitem{LASS}
D.~Aston {\it et al.},
\npb{296}, 493 (1988)

\bibitem{billnote} 
W.M.~Dunwoodie,
Private Communication.



\bibitem{rhopi}
\babar\ Collaboration, B.~Aubert {\it et al.},
\jprl{93}, 051802 (2004)

\end{thebibliography}
